\documentclass[preprint,showpacs,amsmath,amssymb,superscriptaddress,byrevtex,prb]{revtex4}
\usepackage{graphicx}
\usepackage{dcolumn}
\usepackage{bm}
\usepackage[letterpaper,left=20mm,right=20mm,top=4cm,bottom=2cm]{geometry}
\usepackage{amsmath}
\begin{document}
\title{Optical response of strongly coupled metal nanoparticles in
 dimer arrays}

\author{J. J. Xiao}
\affiliation{Department of Physics, The Chinese University of Hong
Kong,
 Shatin, New Territories, Hong Kong}
\author{J. P. Huang}
\affiliation {Department of Physics, The Chinese University of
Hong Kong, Shatin, New Territories, Hong Kong} \affiliation {Max
Planck Institute for Polymer Research, Ackermannweg 10, 55128
 Mainz, Germany}
\author{K. W. Yu}
\email{kwyu@phy.cuhk.edu.hk} \affiliation {Department of Physics,
The Chinese University of Hong Kong,
 Shatin, New Territories, Hong Kong}

\date{\today}

\begin{abstract}
The optical responses of structured array of noble-metal
nanoparticle dimers immersed in a glass matrix are investigated
theoretically, motivated by the recent experimental observation of
the splitting of the surface plasmon bands in silver arrays. To
capture the strong electromagnetic coupling between the two
approaching particles in a silver dimer, the spectral
representation of the multiple image formula has been used, and a
semiclassical description of the silver dielectric function is
adopted from the literature. The splitting of plasmon resonance
band of the incident longitudinal and transverse polarized light
is found to be strongly dependent on the particle diameter and
their separation. Our results are shown in accord with the recent
experimental observation. Moreover, a large redshift for the
longitudinal polarization can be reproduced. The reflectivity
spectrum is further calculated for a dilute suspension of dimer
arrays.
\end{abstract}

\maketitle

\section{introduction}
The optical properties of small metal clusters embedded in a
dielectric medium have attracted extensive attention in recent
years. \cite{Kreibig95,VMShalaev,Olivier02,Yu1997pre,Shalaev98}
The studies have been developed into many new applications in
nanostructure enhanced spectroscopies such as surface-enhanced
Raman scattering and single-target molecule detection,
\cite{VMShalaev,Katrin97,Sheldon00} with near-field observation on
length scales smaller than the wavelength of light. It is known
that strong absorption of light occurs at certain frequencies due
to the collective motions of the conduction electrons in metal
called the surface plasmon resonance, as well as to the inter-band
transition of electrons in deeper levels. The plasmon resonant
frequency depends strongly on the size and the geometry of the
particles, \cite{Jin01,Aizpurua03,Olivier02,Mock02jpc} as well as
on the dielectric function of metal and the surrounding medium.
\cite{Moskovits02} The studies are significant theoretically
because these factors lead to characteristic charge confinement
and strong local field enhancement. \cite{YGu02} On the other
hand, these properties are also of practical importance in the
context of future electronic and optical device applications.

For isolated spherical particles with a diameter much smaller than
the wavelength of light ($d\ll\lambda$), the interactions between
light and metal lead to an homogeneous but oscillating
polarization in the nanoparticles, resulting in a resonant
absorption peak in the visible part of the optical spectrum. The
plasmon resonances in more complex structures of nanoparticles
such as concentric spherical particles, which are the spherical
analog of planar multilayer heterostructures, can be more
complicated. These resonances are due to the hybridization of free
plasmons, which can be pronounced depending on the geometry of the
particles. \cite{Halas02, kwyu97, Oldenburg99} For nanoparticle
ensembles like metal clusters, however, the electromagnetic
coupling between neighboring particles shifts the plasmon
absorption bands. \cite{Ausloos80, Ausloos82} For instance, a
nanoparticle chain can be utilized for building optical waveguides
in the nanoscale. \cite{Stefan02, Pieter03, Romain04} These
structures exhibit lateral mode confinement smaller than the
optical diffraction limit, which cannot be achieved with
conventional waveguides nor with other novel technologies such as
photonic crystals or plasmonic stripe waveguides.

In the linear arrays of nanoparticles, the optical response can
generally be anisotropic, because the interparticle coupling
depends on whether incident light is polarized longitudinal or
transverse to the chain axis. This is also one of the underlying
principles of optically dichroic glass. Nowadays, structured
nanoparticle array can be easily made by electron-beam lithography
\cite{Stefan02prb} or other fabrication techniques.
\cite{Penninkhofa03} On the theoretical side, finite-difference
time-domain (FDTD) simulation methods can accurately reproduce the
spectroscopic properties of plasmon waveguides and optical pulse
propagation in these structures as well. \cite{Pieter03}
Nevertheless, theoretical investigations by the full solution of
Maxwell's equations are complicated due to the coupling between
plasmons of different modes. Although there are already some
fruitful discussions, \cite{Halas97, Stroud04} it is intrinsically
a many-body interaction problem. Generally, two types of
electromagnetic interactions between particles can be
distinguished: near-field coupling and far-field dipolar
interaction depending on the range of interactions concerned. For
particle separation $r$ larger than the wavelength of light, the
far-field dipolar interactions with a $r^{-1}$ dependence
dominate. Much work has focused on these far-field interactions
between metal nanoparticles and their possible applications in
optoelectronic materials. However, relatively little is known
about the nature and the properties of the near-field interactions
of closely spaced metal nanoparticles, which is the object of the
present work. The present work is related to earlier studies of
FDTD by Oliva and Gray,\cite{Oliva03} experiment and simulation by
Su \textit{et al}.,\cite{Su03} the finite elements method by
Kottmann and Martin,\cite{Martin01} as well as the pioneering
works of the discrete dipole approximation (DDA) method by Hao and
Schatza \cite{Schatz04}, who all discovered that the interparticle
spacing in the particle dimers is crucial to their properties.

In this paper, we will use the multiple-image method
\cite{YuCPC00} and Bergman-Milton spectral representation
\cite{Bergman,Bergman79} for a dimer of two approaching particles
\cite{JPHuang02} to investigate the optical extinction and
reflectance spectrum of structured arrays of noble-metal
nanoparticles, motivated by the recent experimental observation of
the splitting of the surface plasmon resonance bands for
polarizations in the arrays. \cite{Stefan02prb, Penninkhofa03} By
taking into account the strong coupling of two approaching
particles in a dimer, we will show that the redshift as large as
$1.5\,$eV into the near-infrared regime observed in experiment
\cite{Penninkhofa03} can be understood. The resonant peak
broadening and splitting for different polarizations can be
predicted from our calculations, and the results for the
reflectance spectrum are also presented.

The rest of the paper is organized as follows. In the next
section, we review the general spectral representation of two
approaching particles. In Sec.~\ref{sec:results}, we examine the
normal-incidence extinction spectrum and reflectance spectrum of a
silver dimer array immersed in a glass matrix, which is followed
by discussion and conclusion in Sec.~\ref{sec:discussion}

\section{Formalism}
\label{sec:theory} First we review some formulae similar to those
appeared in Ref.$\,$32, however, in a much different context.
Considering an isolated spherical metal particle in a lossless
isotropic dielectric matrix with real permittivity
$\varepsilon_m$, the complex permittivity of the metal particle is
$\tilde\varepsilon(\omega)$, where $\omega$ is the frequency of
the external electric field $\vec{E}_0$, and will be discussed in
detail afterwards. In this case, the dipole moment induced inside
the particle is
\begin{equation}\label{eqs:singledipole}
\tilde p=\frac{1}{8}\varepsilon_m \tilde{\beta}d^3E_0,
\end{equation}
where
\begin{equation}\label{eqs:singldipolefactor}
\tilde{\beta}=\frac{\tilde\varepsilon(\omega)-\varepsilon_m}{\tilde\varepsilon(\omega)+2\varepsilon_m}
\end{equation}
is the dipole factor, which defines the polarizability of the
particle against the host and is related to extinction coefficient
directly, $d$ is the diameter of the particle. To account for the
multipolar interaction between a pair of particles (\textit{i.e.,}
a dimer) with spacing $\sigma$ (center-to-center distance
$r=\sigma+d$), we use the multiple image formula. \cite{YuCPC00}
When the dimer is subjected to an unpolarized field, the average
of the total dipole moment of one particle is given by
\begin{equation}\label{eqs:totaldipole}
    \tilde{p}^*=\tilde{p}_T\langle\cos ^2\theta\rangle+\tilde{p}_L\langle\sin ^2\theta\rangle
 =\frac{1}{2}(\tilde{p}_T+\tilde{p}_L),
\end{equation}
where $\theta$ is the angle between the dipole moment and the line
joining the centers of the two particles. Here $\tilde{p}_L$ and
$\tilde{p}_T$ are the longitudinal and transverse dipole moment,
respectively, \cite{JPHuang02}
\begin{eqnarray}
\label{dmoment} \tilde{p}_L&=&\tilde{p}\sum_{n=0}^\infty(
2\tilde\beta)^n \left(\frac{\sinh \alpha} {\sinh (n+1)\alpha}\right)^3,\nonumber\\
\tilde{p}_T&=&\tilde{p}\sum_{n=0}^\infty(-\tilde\beta)^n
\left(\frac{\sinh \alpha}
 {\sinh (n+1)\alpha}\right)^3,
\end{eqnarray}
where $\alpha$ satisfies the relation $\cosh \alpha=(\sigma+d)/d$.
Now the new dipole factors ($\tilde{\beta}_L$ and
$\tilde{\beta}_T$) of a particle in the dimer can be extracted for
the longitudinal and transverse field case, respectively. Using
the spectral representation, \cite{Bergman} we have
\begin{eqnarray}
\label{eqs:factor}
\tilde\beta_L=\sum_{n=1}^{\infty}\frac{F^{(L)}_{n}}{\tilde{s}-s^{(L)}_{n}},\nonumber\\
\tilde
\beta_T=\sum_{n=1}^{\infty}\frac{F^{(T)}_{n}}{\tilde{s}-s^{(T)}_{n}},
\end{eqnarray}
with the complex material parameter
\begin{equation}\label{eqs:parame}
\tilde{s}=\left(1-\frac{\tilde\varepsilon(\omega)}{\varepsilon_m}\right)^{-1},
\end{equation}
where
\begin{eqnarray}
F^{(L)}_{n}=F^{(T)}_{n}=-\frac{4}{3}n(n+1)\sinh ^3\alpha
 e^{-(2n+1)\alpha},\nonumber\\
s^{(L)}_{n}=\frac{1}{3}[1-2e^{-(1+2n)\alpha}], \
s^{(T)}_{n}=\frac{1}{3}[1+e^{-(1+2n)\alpha}].
\end{eqnarray}
In case of unpolarized field, the averaged dipole factor
$\tilde{\beta}^*$ can be derived directly from
Eq.~(\ref{eqs:totaldipole}), Eq.~(\ref{dmoment}) and
Eq.~(\ref{eqs:factor}), namely,
\begin{equation}\label{eqs:avefactor}
\tilde{\beta}^*=\frac{1}{2}\sum_{n=1}^{\infty}
\left(\frac{F^{(L)}_{n}}{\tilde{s}-s^{(L)}_{n}}+\frac{F^{(T)}_{n}}{\tilde{s}-s^{(T)}_{n}}\right).
\end{equation}
Eq.~(\ref{eqs:factor}) (or Eq.~(\ref{eqs:avefactor})) is an exact
transformation of the multiple image expression, \cite{YuCPC00}
and consists of a set of discrete poles $s^{(L)}_n$ and
$s^{(T)}_n$, which deviates from $1/3$ (pole of an isolated
spherical particle). In particular, the longitudinal and
transverse poles $s^{(L)}_n$ and $s^{(T)}_n$ shift asymmetrically
to different sides from $1/3$. That is, an unpolarized field can
excite all poles at both sides. The poles almost collapse to that
of an isolated sphere ($s^{(L)}_n$ and $s^{(T)}_n\rightarrow1/3$)
if $\sigma\geqslant d$, indicating that the multipolar interaction
is negligible. However, when the two particles approach to each
other and finally touch, the longitudinal (transverse) poles
decrease (increase) far from $1/3$. Thus, in this case, one should
take into account the effect of multipolar interactions [see Fig.
5 in Ref.~32 for details].

The complex dielectric function $\varepsilon(\omega)$ is crucial
to the optical properties of metal-dielectric systems.
\cite{Kreibig95} For noble metals, it can generally be described
by the free electron Drude-Lorentz-Sommerfeld model plus an
additive complex contribution from interband transition,
\textit{i.e.,}
$\varepsilon(\omega)=1+\chi^{DS}(\omega)+\chi^{IB}(\omega)$. A
complicated function of the dielectric dispersion of Ag takes the
form
\begin{equation}\label{eqs:lorentz}
\varepsilon (\omega ) = 1 + \varepsilon _\infty - \frac{\omega
_p^2 }{\omega ^2 + i\omega \gamma } + \sum\limits_j^N {\frac{a_j
}{\omega _{oj}^2 - \omega ^2 - i\omega \Gamma _j }},
\end{equation}
which could be adopted to approximate the measured dielectric
function over a wide wavelength range. \cite{Moskovits02} In
Eq.~(\ref{eqs:lorentz}), $a_j$ may be negative. The sum over $N$
Lorentz functions and the constants are meant to reproduce the
interband and all other non-Drude contributions to the dielectric
function. Lorentz functions are chosen because it is known that
$\varepsilon(\omega )$ must obey the Kramers--Kronig relations.
However, in the frequency range of interest ($1\sim4.5\,$ev),
\cite{Penninkhofa03} a modified Drude model is easier to describe
the dielectric response of Ag: \cite{Pieter04}
\begin{equation}\label{eqs:drude}
\varepsilon (\omega ) = \varepsilon _h - \frac{(\varepsilon _s -
\varepsilon _h )\omega _p^2 }{\omega(\omega + i\gamma )},
\end{equation}
with plasmon resonant frequency $\omega _p = 1.72\times
10^{16}\,$rad/s and with the high-frequency limit dielectric
function $\varepsilon _h = 5.45$, static dielectric function
$\varepsilon _s = 6.18$. These values were fitted out to be in
good correspondence with the literature values. \cite{Johnson72,
Moskovits02} And the collision frequency $\gamma$  in the material
is assumed to be related to the particle diameter $d$ around
$10\,$nm by \cite{Peters00}
\begin{equation}
\gamma=\frac{\nu_f}{\ell}+\frac{2\nu_f}{d},
\label{eqs:linewidth}
\end{equation}
with bulk Fermi velocity  $\nu_f=1.38\times10^6\,$m/s, room
temperature electron mean free path $\ell=52\,$nm. For $d=10\,$nm,
$\gamma=3.025\times10^{14}$; for $d=5\,$nm,
$\gamma=5.785\times10^{14}$. These results show that the mean free
path of an electron in a nanoparticle is reduced compared to its
bulk value due to inelastic collisions with the particle surface.
The $\gamma$ values are taken in our latter calculations of
different metal-dielectric systems, and compared to the two
experimental samples, within which the diameters of Ag
nanoparticles were in the span of $5\sim15\,$nm. \cite{Peters00,
Penninkhofa03} Note that $\gamma$ determines the linewidth of the
resonant peak. In the diameter range under consideration,
Eq.~(\ref{eqs:linewidth}) is safe \cite{Molina02} and indicates
that a smaller particle diameter $d$ leads to a wider resonant
peak. We are not intended to quantitatively compare with the
experimental data of Ref. 22, otherwise, we would be restricted to
a somewhat more rigid size-dependent dielectric function, for
example, as Westcott \textit{et al}.\cite{Halas02} Nevertheless,
the local dielectric treatment is satisfactory as Hao and Schatz
pointed out that the significant effects of size-dependent
dielectric responses come to appear for particles with diameter less
than $5\,$nm.\cite{Schatz04}

Let us use $\varepsilon_1(\omega)$ and $\varepsilon_2(\omega)$ to
denote the real and imaginary part of the dielectric function
obtained by Eq.~(\ref{eqs:drude}), respectively, that is
$\varepsilon_(\omega)=\varepsilon_1(\omega)+i\varepsilon_2(\omega)$.
Fig.~\ref{fig:drude} shows $\varepsilon_1(\omega)$ and
$\varepsilon_2(\omega)$ versus light wavelength $\lambda$ in the
span of $250\sim1500\,$nm (\textit{i.e.} photon energy around
$0.8\sim5\,$eV). A negative $\varepsilon_1(\omega)$ is guaranteed
for the proper phase relation between the external field and
particle polarization. The dielectric function changes slightly
when the particle diameter decreases from $10\,$nm to $5\,$nm [not
shown in Fig.~\ref{fig:drude}], however, the resonant line shape
is very sensitive to $\gamma$, \textit{i.e.} the particle diameter
[see Sec.~\ref{sec:results}]. $d\ll\lambda$ ensures that the
plasma resonance is in \textit{quasi-static regime}, so phase
retardation is negligible, effects of higher multipoles can also
be neglected for isolated spherical particle, which means that
dipole plasmon resonance dominates. \cite{Kreibig95}

\section{Numerical Results}
\label{sec:results}

Now we consider an array of silver dimer immersed in a glass
matrix of refractive index 1.61, with the spacing between the two
particles in a dimer being less than their diameter
($\sigma\leqslant d$). Any two dimers are assumed to be far away
enough, so the dimer-dimer interaction can simply be at far-field
approximation, which is neglected in our calculation for
simplicity. In the particle diameter regime around $10\,$nm,
dipole absorption contribution dominates the scattering effect,
although dipole scattering increases and dipole absorption fades
away for increasing particle sizes. \cite{Kreibig95} So in the
\textit{quasi-static regime}, the extinction coefficient of a
well-dispersed collection of small particles is mainly contributed
by absorption, with absorption cross section proportional to
$\omega$Im($\tilde{\beta}$). \cite{Angel93, Kreibig95} The complex
value expressions of dipole factor in
Eq.~(\ref{eqs:singldipolefactor}), Eq.~(\ref{eqs:factor}) and
Eq.~(\ref{eqs:avefactor}) lead to different resonant peaks at
different frequencies. To calculate optical extinction,
$\tilde\beta$ is taken for well-dispersed (isolated)
nanoparticles, while $\tilde\beta_L$ ($\tilde\beta_T$) is adopted
for an array of dimers.

Figure~\ref{fig:smp1_abso} shows the optical extinction spectra of
an array of dimers with particle diameter $d=5\,$nm, spacing
$\sigma$ is $0.5\,$nm and $1.5\,$nm, respectively. For comparison,
solid curve is plotted for the extinction spectrum of isolated
silver particles. The surface plasmon resonant peak is located
around $410\,$nm ($\sim3.0\,$eV), which is in agreement with the
first ion-exchanged sample irradiated by $1\,$Mev Xe in Ref.~22.
The sample contains randomly dispersed silver nanocrystals and the
resonant band is polarization independent. Long-dashed curves and
medium-dashed curves in Fig.~\ref{fig:smp1_abso} are the
extinction spectra for the array of dimers with illumination light
polarized in the longitudinal and transverse direction,
respectively. It is clear that the plasmon resonance band for
longitudinal polarization is redshifted with respect to that of
isolated particles, whereas the plasmon resonance band for
transverse polarization is blueshifted. These were also observed
in the experiment. \cite{Penninkhofa03} In detail, after the
sample was subsequently irradiated by $30\,$Mev Si with fluences
up to $2\times10^{14}$/cm$^{2}$, clear alignment of Ag
nanocrystals was observed along the ion-beam direction.
Additionally, farther redshift and blueshift occur when decreasing
the spacing of the two nanoparticles in a dimer from $1.5\,$nm to
$0.5\,$nm [see Fig.~\ref{fig:smp1_abso}]. That is, a stronger
electromagnetic coupling induces further band shifts. \cite{YGu02}

Similarly, an array of dimer with a larger particle diameter
$d=10\,$nm are investigated in Fig.~\ref{fig:smp2_abso}, for
different spacing $\sigma=0.5$, $1.5$, $2.5$ and $3.5\,$nm.
Splitting of the resonant peak for both  the longitudinal and
transverse polarized light can also be observed. However, a second
resonant band appears as the two particles in the dimer approach
to each other. The second peak position is around $830\,$nm (close
to $1.5\,$eV) when spacing decreases to $\sigma=0.5\,$nm [see
Fig.~\ref{fig:smp2_abso}(a)], this is in good agreement with the
experimental observation of the second sample in Ref.~22. For this
sample, growing and more compact alignment of the silver
nanocrystals  are assumed to happen in response to higher Si ion
fluences irradiation (up to $1\times10^{15}$/cm$^2$), and much
higher irradiation fluences induces much larger splitting of the
resonant band for both the longitudinal and transverse
polarizations. In fact, all of these are also obtained from
Fig.~\ref{fig:smp2_abso}. In principle, even in the quasi-static
regime, there are different causes of multipeak behavior of
optical spectra: (1) the splitting of the dipole mode owing to
nonspherical particle shapes, (2) the excitation of higher
multipole modes in irregularly shaped clusters as a result of
inhomogeneous polarization (In this case, number of resonances
strongly increases when the section symmetry decreases), (3) the
enhanced excitation of multipoles due to image interactions for
spheres. Multipeak structures can also be produced by appropriate
$\varepsilon_1(\omega)$ spectra, and may be damped away if
$\varepsilon_2(\omega)$ is sufficiently large. But as
Fig.\ref{fig:drude} shows, the model dielectric function resulting
from experiments of silver nanoparticles doesn't exhibit these
behaviors, so we are confident that the multipeak behavior
observed in the sample is substantially due to the strongly
coupling between the two particles in a dimer (\textit{i.e.} (3)),
because no obvious identical irregular shape can be seen for the
samples. \cite{Penninkhofa03}

The largest shift of resonant peak of the dimer array is shown for
the longitudinal and transverse field cases in
Fig.~\ref{fig:splitting}, at $d=10\,$nm. The figure shows the
farthest shifting of the peak positions versus spacing $\sigma$.
The results are obtained by first calculating the extinction
spectra with different $\sigma$ in the range of $0.5\sim20\,$nm,
and then finding out the position of resonant peak at the longest
(shortest) wavelength for the longitudinal (transverse)
polarization. Large shifting is obvious only when $\sigma/d<1$,
and this can also be understood from the spectral representation
in the insert of Fig.~\ref{fig:abso_ave}(b), which demonstrates
that the spectral poles collapse to $1/3$ when $\sigma$ tends to
be larger than $10\,$nm. Within the spacing of $\sigma<d$, the
redshift for the longitudinal polarization (diamond) is obviously
stronger than the blueshift for transverse polarization (circle).
The large splitting is due to both the dipole mode coupling
(collective excitation mode) and the excitation of multipole modes
by image interactions.

We also note that the main resonant peak is narrower in
Fig.~\ref{fig:smp2_abso} than that in Fig.~\ref{fig:smp1_abso},
which is mainly ascribed to the different intrinsic damping
efficient concerning the particle diameter. The narrowing of the
resonant peak for increasing particle diameter was also observed
in the experiments. \cite{Penninkhofa03} Note that no obvious peak
shifting is observed in the experiments for isolated particles
after their growing.

It is instructive to investigate the optical extinction properties
of the array of dimers for an unpolarized light. We take the
average of the longitudinal and transverse dipole factor,
\textit{i.e.} Eq.~(\ref{eqs:avefactor}) to obtain the extinction
spectra. Results are shown in Fig.~\ref{fig:abso_ave} for the two
different cases of particle diameter $d=5\,$nm and $10\,$nm,
respectively. Corresponding poles and residues of $n=1$ to 10 in
the spectral representation are given in the inserts. It can be
clearly seen that extinction spectra still change when particles
aggregate into structured array from a well-dispersed assemble,
even though the illumination light keeps unpolarized. This is due
to the asymmetric coupling in different topology of the field
distribution inside and in the vicinity of the particles.
\cite{YGu02} Note that the shifting tendency of plasmon-resonance
band could be related to the pole deviating from $1/3$. For
specific spacing, say, $\sigma=0.5\,$nm, the pole of $n=1$ of the
dimer with large particles [see insert in
Fig.~\ref{fig:abso_ave}(b)] is further away from 1/3 than that
with small particles [see insert in Fig.~\ref{fig:abso_ave}(a)].
Equivalently, there is a corresponding resonant peak appearing
[Fig.~\ref{fig:abso_ave}(b)] at long wavelength regime for the
case of large particles ($10\,$nm), whereas no obvious peak
appearing [Fig.~\ref{fig:abso_ave}(a)] for the case of small
particles ($5\,$nm). Theoretically, the discrete terms in the
spectral representation should generate a series of resonant
bands, but they are generally very close and superpose to each
other, so it is hard to resolve them. In many cases the effect is
broadening of the resonance only, so the plasmon modes remain
spectrally unseparated.

For interest, we compare the reflectivity spectra for the
different cases. In the dilute limit, the effective dielectric
function of the composite is given by \cite{kwyu93, Yuen97}
\begin{equation}
\label{eqs:reflect}
\tilde\varepsilon_e=\varepsilon_m+3\varepsilon_m p \tilde\beta,
\end{equation}
where p is the volume fraction of the silver particles.
Reflectance at normal incidence is
$R=|(1-\sqrt{\tilde\varepsilon_e})/(1+\sqrt{\tilde\varepsilon_e})|^2$
.\cite{kwyu97} Fig.~\ref{fig:smp1_ref} and Fig.~\ref{fig:smp2_ref}
are plotted as the reflectance versus photon energy of the
illumination light for the different arrays discussed above, with
particle diameter $d=5\,$ and $d=10\,$nm respectively. Volume
fraction $p=0.1$ [see Fig.~\ref{fig:smp1_ref}(a), (b) and
Fig.~\ref{fig:smp2_ref}(a), (b)] and $p=0.01$ [see
Fig.~\ref{fig:smp1_ref}(c), (d) and Fig.~\ref{fig:smp2_ref}(c),
(d)] are taken in the calculations. We can see slight shift of the
reflectivity spectra for light with longitudinal and transverse
polarization as compared to the case of isolated particles [solid
curves in Fig.~\ref{fig:smp1_ref} and Fig.~\ref{fig:smp2_ref}].
Reflectance decreasing in low-frequency regime is notable when
spacing decreases [see Fig.~\ref{fig:smp1_ref}(a), (c) and
Fig.~\ref{fig:smp2_ref}(a), (c)]. A large fluctuation in response
to longitudinally polarized light is observed in the case of the
larger particles with a small spacing [long-dashed curves in
Fig.~\ref{fig:smp2_ref}]. Meanwhile, increasing the volume
fraction of the metal nanoparticles leads to the descending main
peak of the reflectivity spectra. But note that the order of the
absolute value of reflectance is different for the two different
volume fraction. These are also ascribed to the very strong
electromagnetic coupling between the nearby particles and the
strongly enhanced local field.

\section{discussion and conclusion}
\label{sec:discussion} Here a few comments on our results are in
order. Firstly, the tail going up of the extinction spectra in
Ref.~22 is ascribed to the absorption of the matrix itself, see
the reference measurement for a Ag ion exchanged sample that was
irradiated with Si only. It does not show a plasmon absorption
band and is colorless, because there were no silver nanocrystals
formed. We didn't take it into account in our calculation of the
extinction spectrum. Furthermore, the experimental spectra are
broader than our results, this is due to particle size
distribution and coupling of particles in a chain through
many-body interaction although we show that two-body interaction
contributes little when particles are far away with spacing larger
than their diameter.

In the present work, we are concerned with the multipolar
interaction between particles, \textit{i.e.}, we focus mainly on
the dimerization effect. In doing so, we could neglect the
non-local effects in the present classical treatment. It is
believed that the significant effects of size-dependent dielectric
responses come to appear for particles with diameter less than 5
nm.\cite{Schatz04}Although near adjacence may enhance nonlocal
contribution, our calculation results depend on the
spacing-diameter ratio $\sigma/d$, rather than the absolute value
of $\sigma$. That is, the spacing in use is scaled by the particle
size, if regardless of the retardation effect. So we also didn't
consider the suppression of the dielectric confinement due to
quantum penetration effects with very small interparticle gaps. We
should worry about retardation effects when the particles size
becomes large.

Due to strong coupling in the dimer, the enhanced excitation of
multipoles of the electric field that occurs in the vicinity of
the dimer appears, which is responsible for the excitation of
spectrally distinct higher order plasmon modes, such as the
quadrupole plasmon-peak feature around 900nm. Crosstalk between
adjacent dimers is expected to have a negligible effect on the
optical spectra since it takes place via far-field scattering of
the individual nanoparticles with a distance dependence of
$r^{-1}$, whereas near-field interactions of adjacent particles in
each dimer show a $r^{-3}$ dependence and dominate at small
distances, which is captured by our multiple image model.

Maier \textit{et al.} used a coupled dipole model in a particle
chain and predicted a value of $2:1$ of the ratio of peak shifts
of the longitudinal collective modes to peak shifts of the
transverse collective modes, which is smaller than experiment
value $2.3:1$. \cite{Stefan02prb} We believe that if one takes
into account the multiple image interactions in the nanoparticle
chain, the theoretically calculated value will be increased. Work
are under progress in this way.

We can take one step forward to include the nonlinear
characteristics of noble metal particles. For instance, based on
Eq.~(\ref{eqs:reflect}), we could derive the effective third-order
nonlinear susceptibility and then the nonlinear enhancement may be
studied by taking into account multipolar interactions. Regarding
nonlinear enhancement due to dimerization, we can formulate some
equations, based on, say, Yuen and Yu. \cite{kwyu97} It is also
interesting to apply the present theory to the polydisperse size
case, in which the two particles have different diameters.

\section{acknowledgement}
This work was supported by the RGC Earmarked Grant.

\newpage
\textbf{\begin{center} {Figure Captions}\end{center}}

\begin{figure}[hbpt!]
\includegraphics[scale=1]{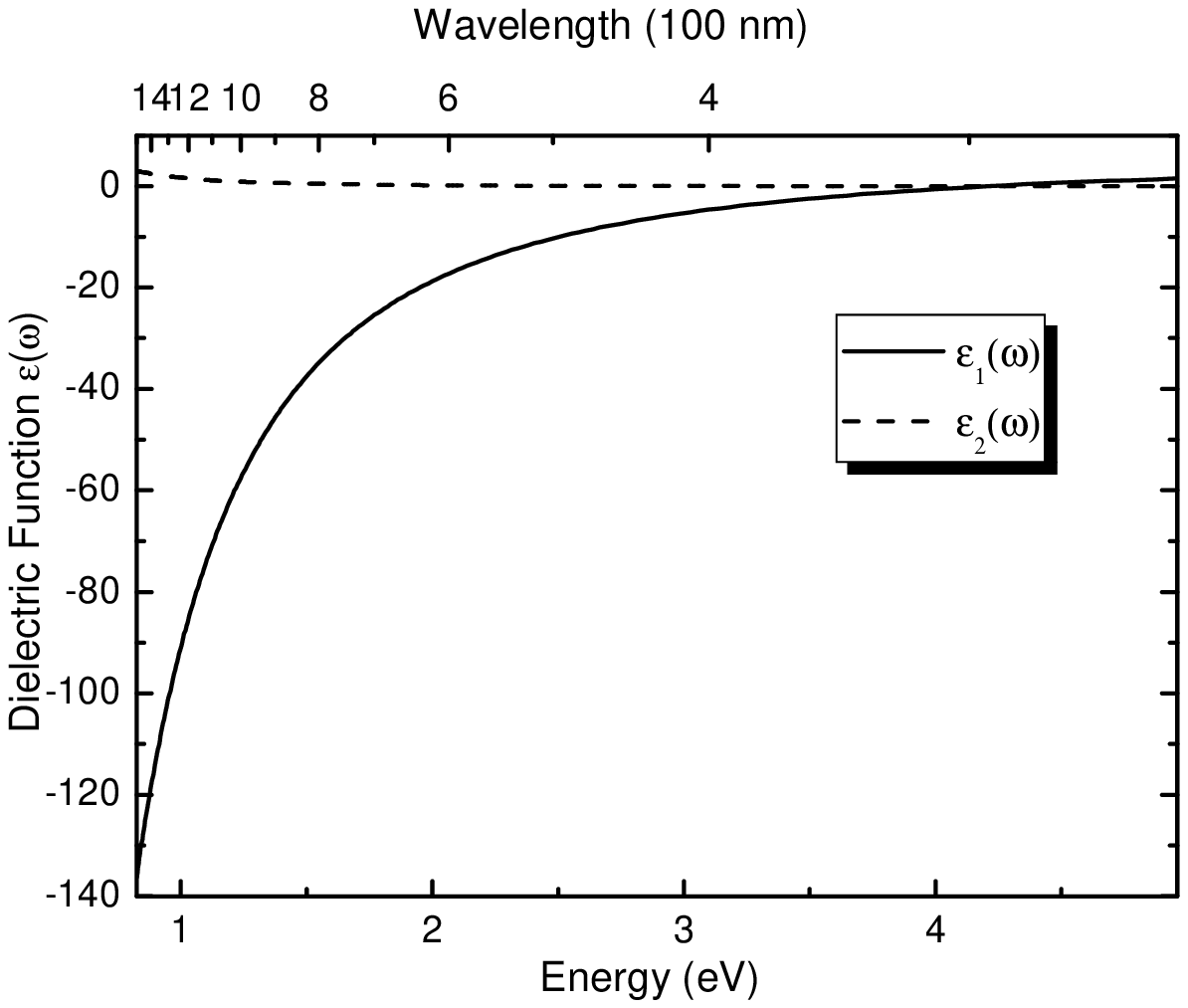}
\caption{} \label{fig:drude}
\end{figure}

\textbf{Figure \ref{fig:drude}}: Complex dielectric function of
silver particles obtained from Eq.~(\ref{eqs:drude}). Parameters:
$d=10\,$nm, others given in the context.

\begin{figure*}[hbpt!]
\begin{center}
\includegraphics[scale=0.5]{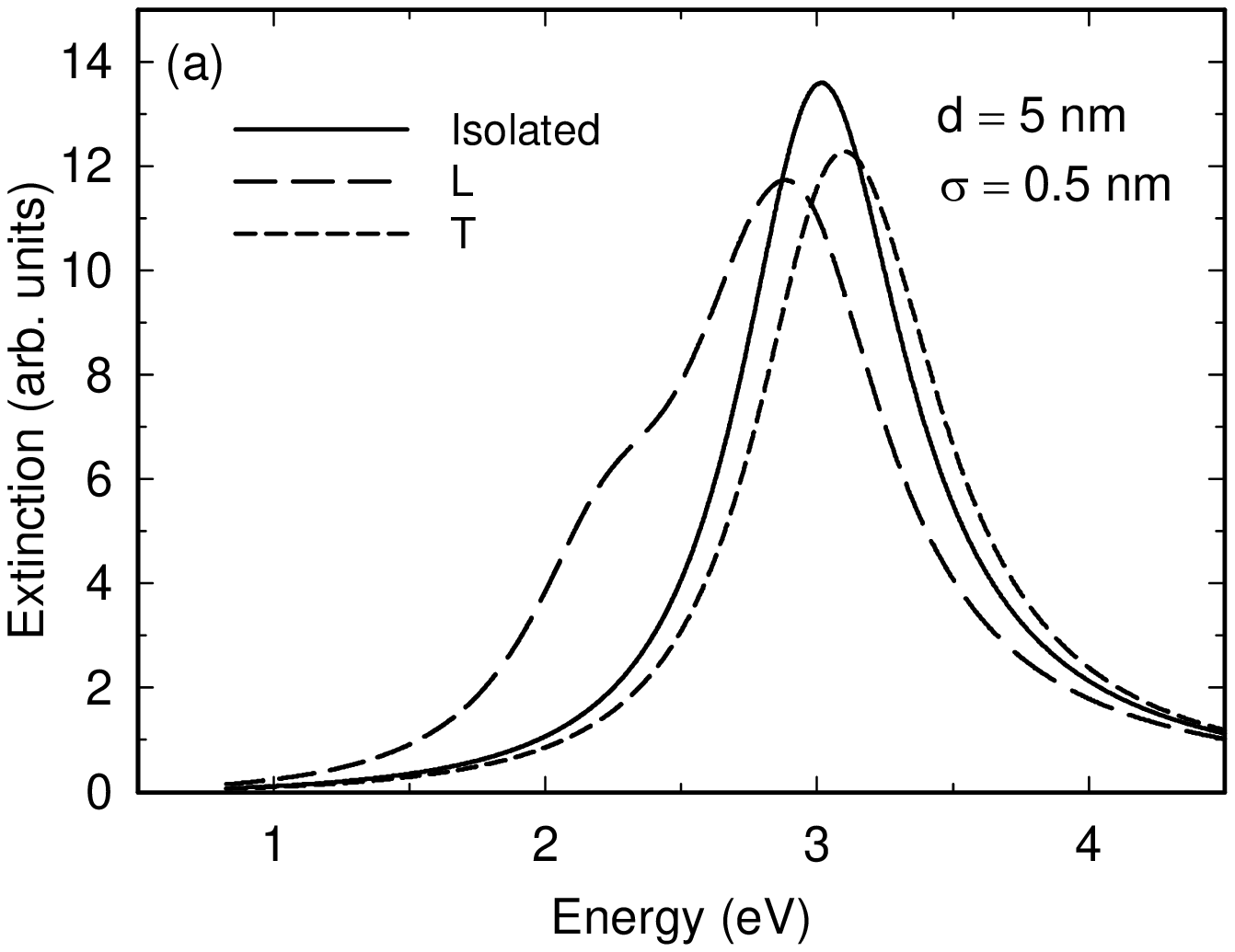}
\includegraphics[scale=0.5,trim=0cm 0cm 0cm -1cm]{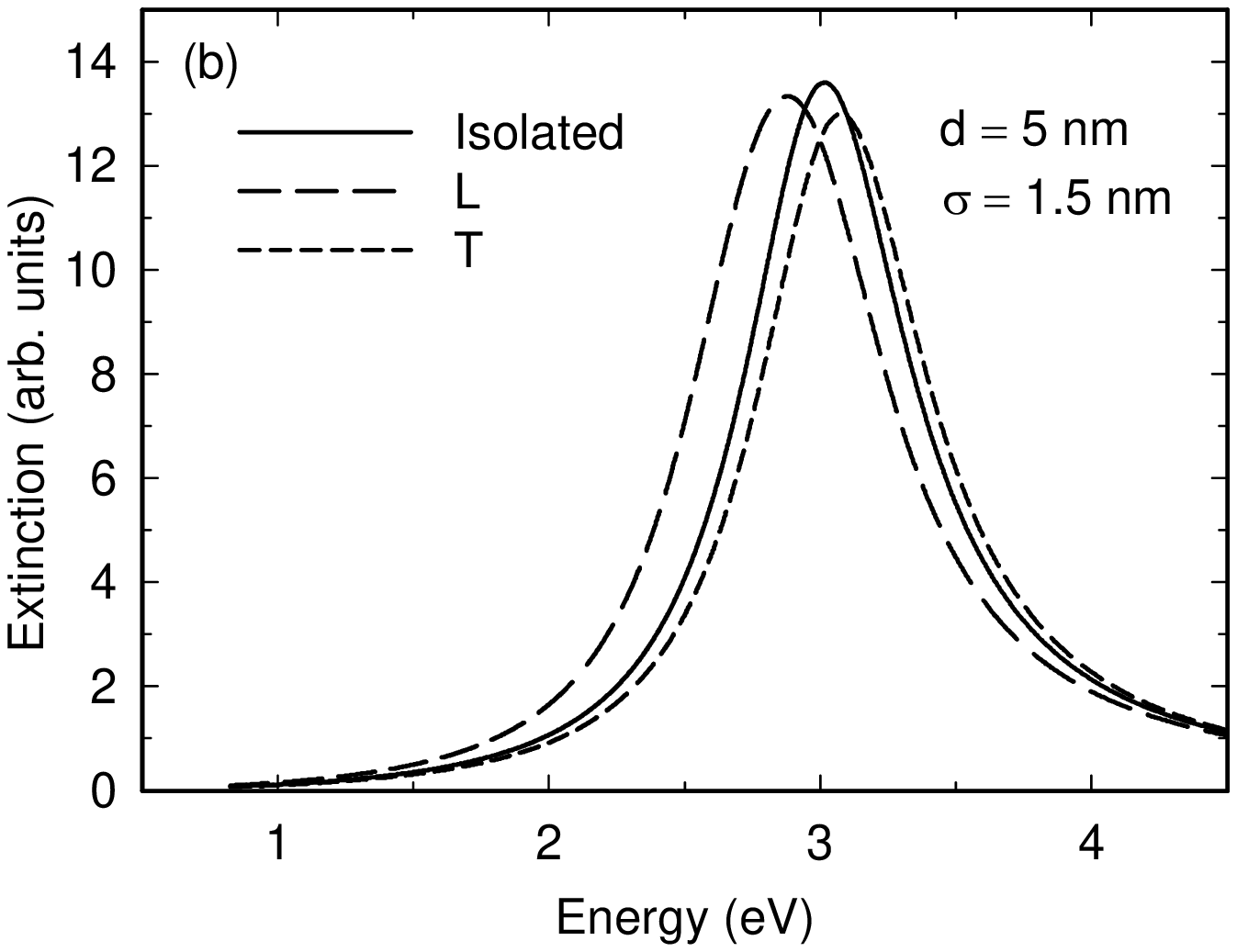}
\end{center}
\caption{} \label{fig:smp1_abso}
\end{figure*}

\textbf{Figure \ref{fig:smp1_abso}}: Extinction spectra for an
array of dimer with particle diameter $d=5\,$nm,  at two different
spacing $\sigma$. The polarization of the incident light is
longitudinal (long-dashed curve) or transverse (medium-dashed
curve) to the axis of the dimer. For reference, solid curve is the
extinction spectra of isolated and well-dispersed particle
collection.

\begin{figure*}[hbpt!]
\begin{center}
\includegraphics[scale=0.5]{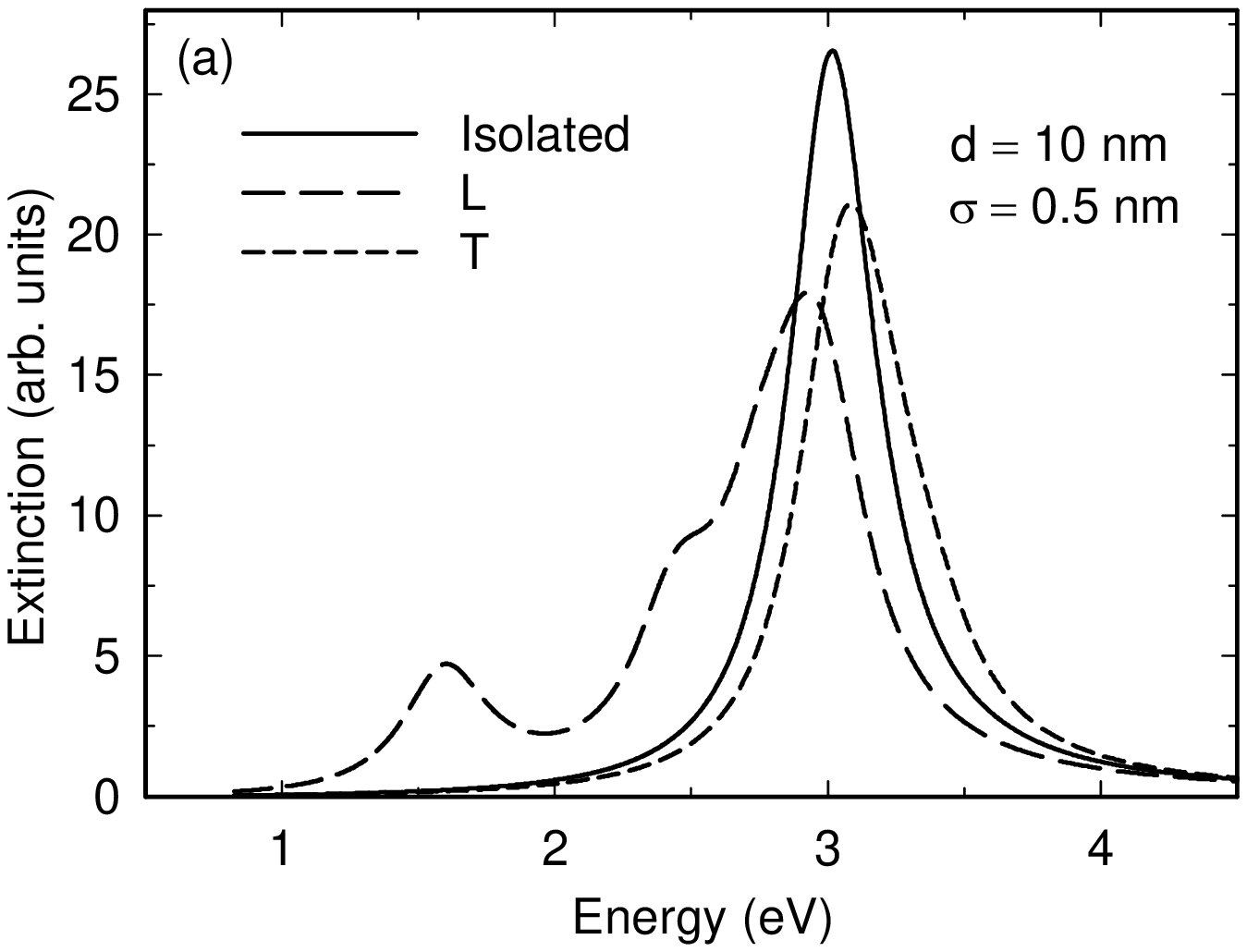}
\includegraphics[scale=0.5]{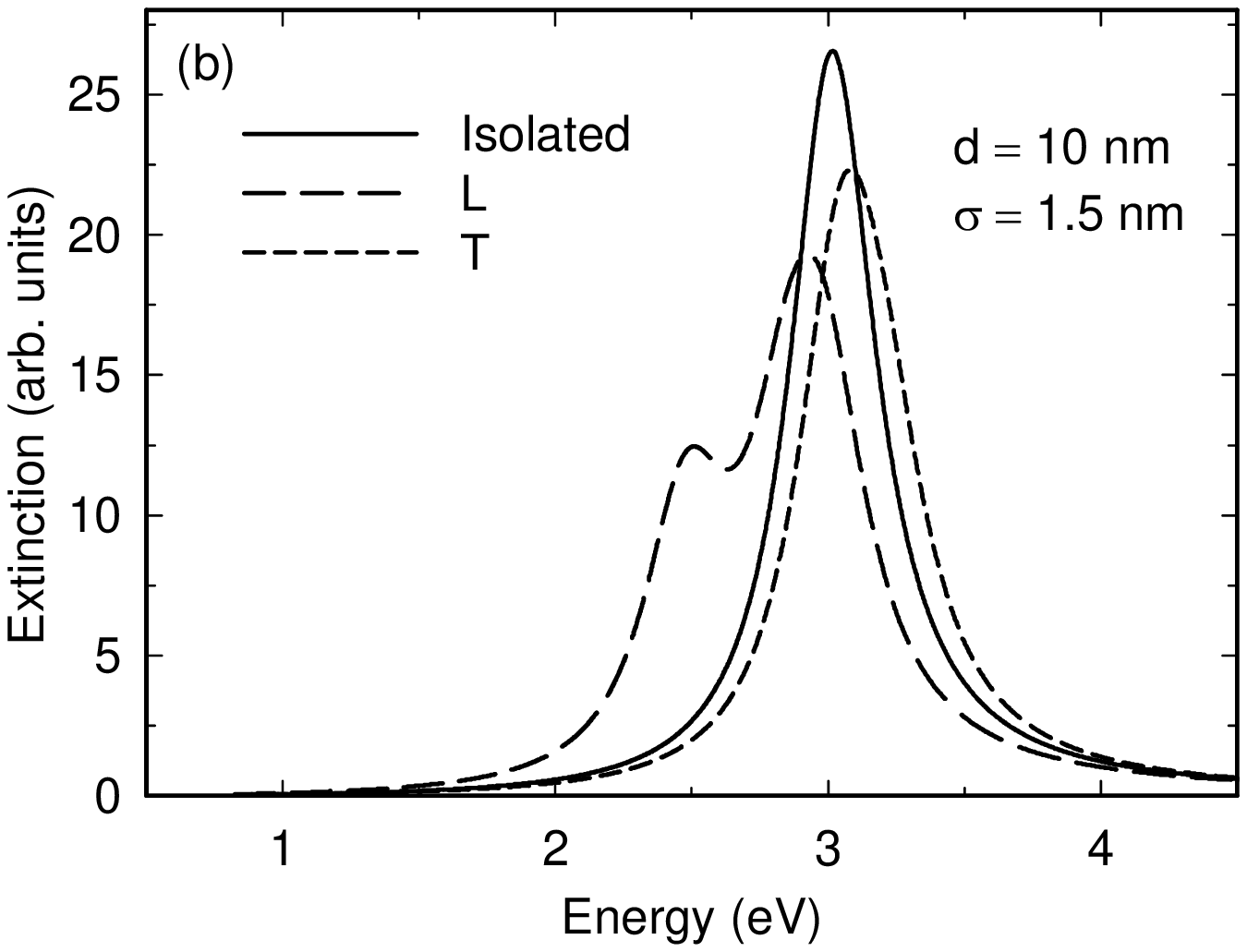}
\includegraphics[scale=0.5,trim=0cm 0cm 0cm -1.2cm]{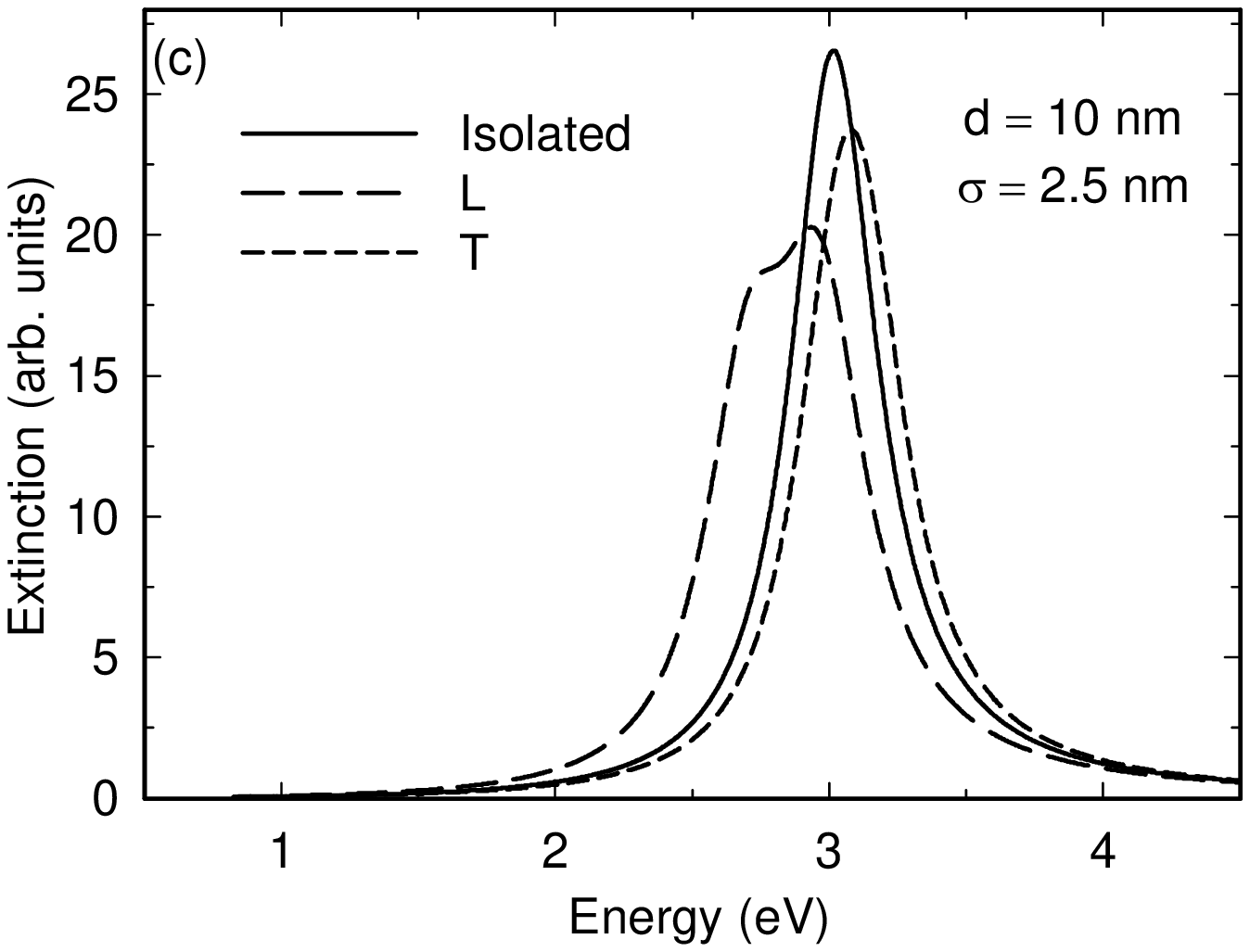}
\includegraphics[scale=0.5,trim=0cm 0cm 0cm -1.2cm]{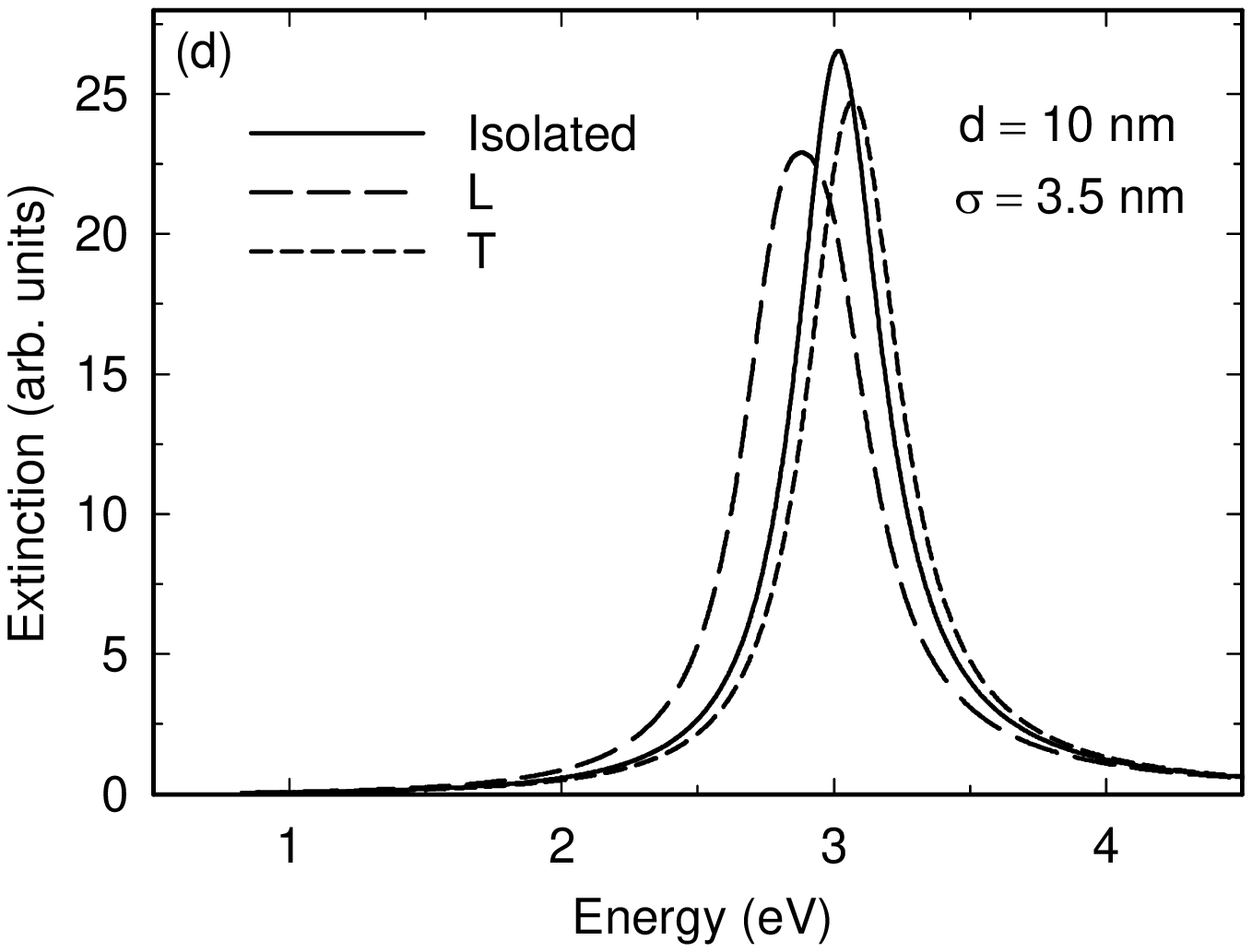}
\end{center}
\caption{}\label{fig:smp2_abso}
\end{figure*}

\textbf{Figure \ref{fig:smp2_abso}}: Splitting of the extinction
spectra for an array of dimer with particle diameter $d=10\,$nm,
at four different spacing $\sigma$. Others the same as in
Fig.~\ref{fig:smp1_abso}.

\begin{figure}[hbpt!]
\begin{center}
\includegraphics[scale=0.7]{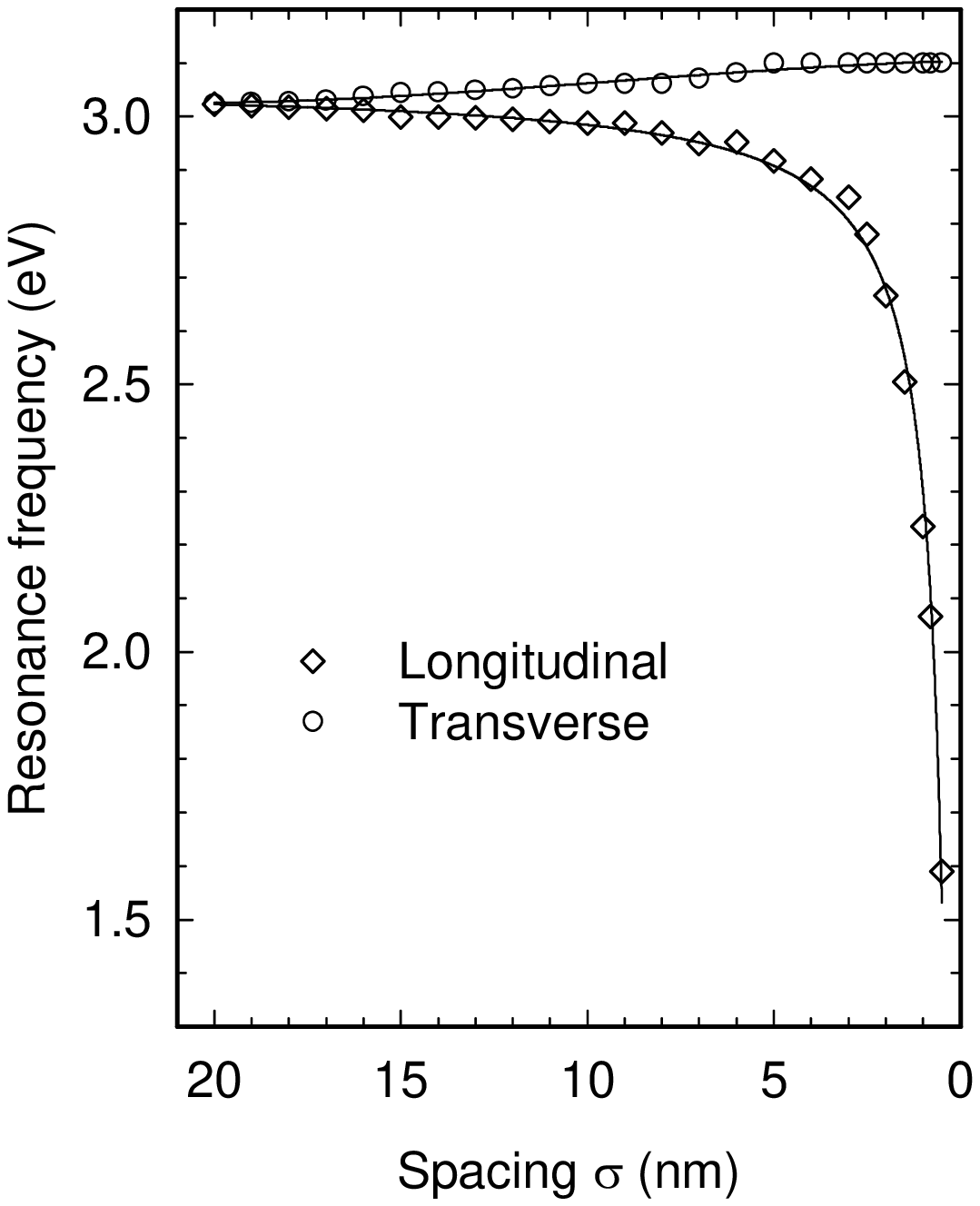}
\end{center}
\caption{}\label{fig:splitting}
\end{figure}

\textbf{Figure \ref{fig:splitting}}: The resonant peak splitting
for an dimer array of diameter $d=10\,$nm as the spacing decreases
from $20\,$nm to $0.5\,$nm. The lines are guides to the eye.

\begin{figure*}[hbpt!]
\begin{center}
\includegraphics[scale=0.5]{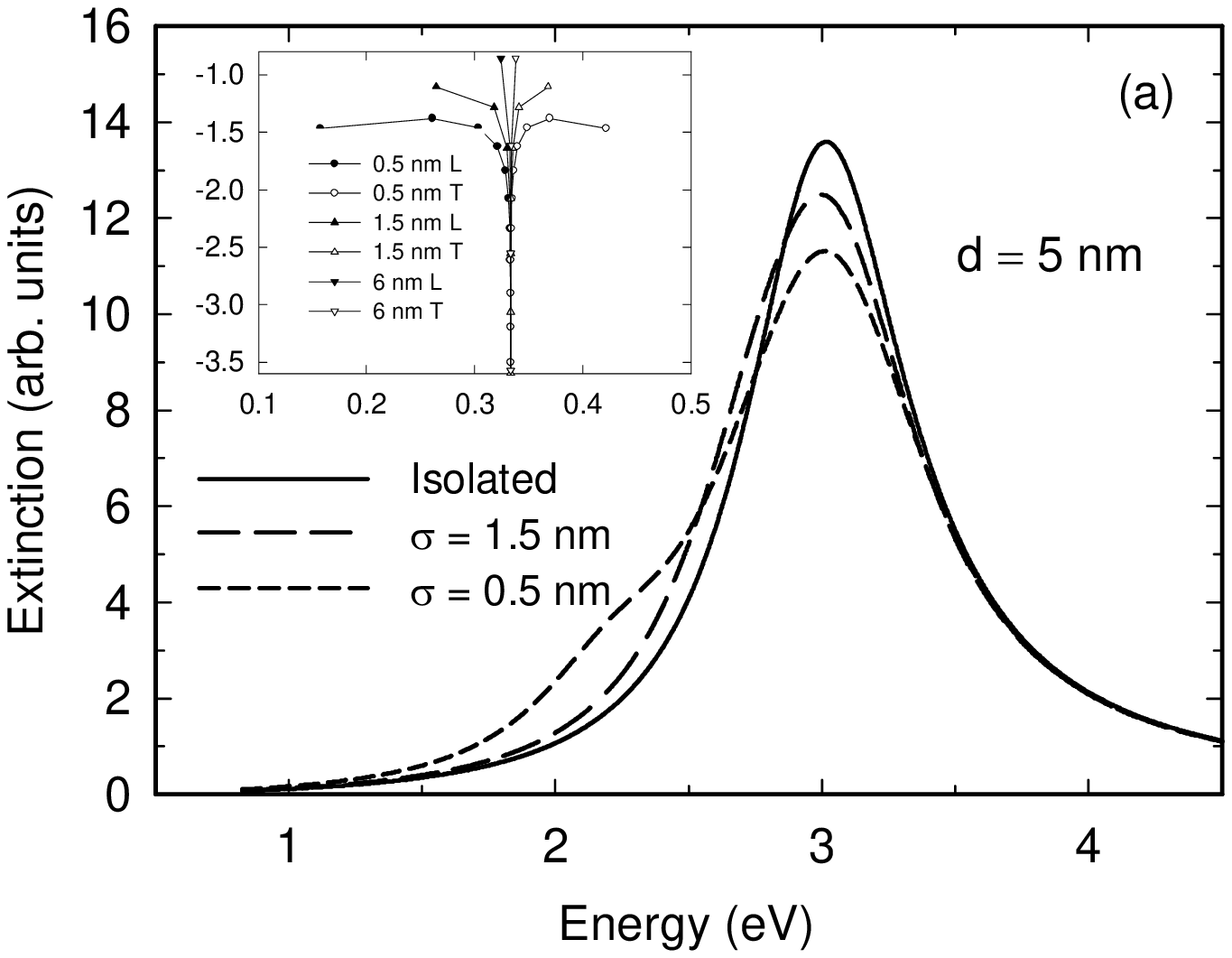}
\includegraphics[scale=0.5,trim=0cm 0cm 0cm -1.8cm]{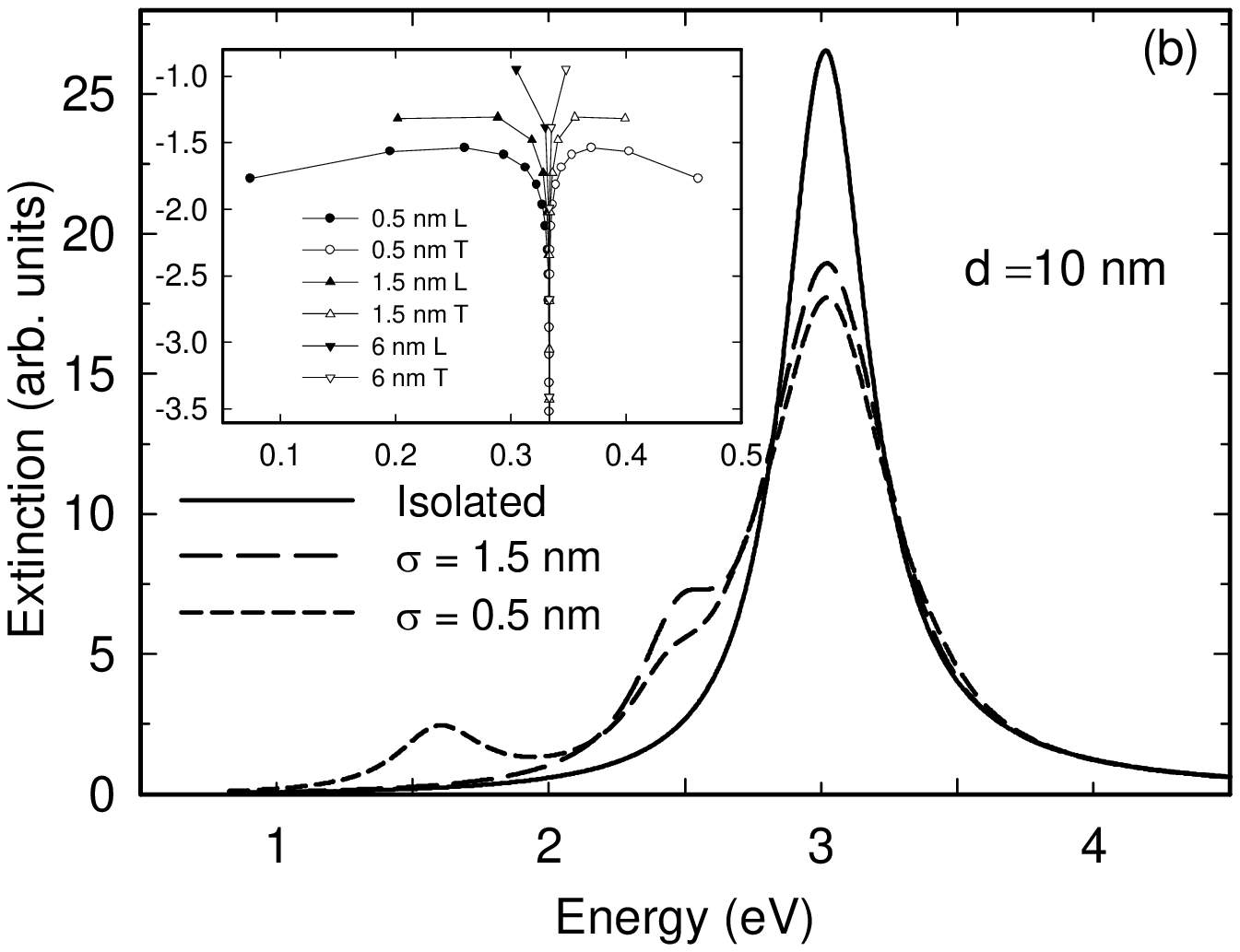}
\end{center}
\caption{}\label{fig:abso_ave}
\end{figure*}

\textbf{Figure \ref{fig:abso_ave}}: Unpolarized light extinction
spectrum for an array of dimer with particle diameter $d=10\,$nm
and $5\,$nm respectively, at different spacing $\sigma$. Inserts
in each panel are the poles and residues of the spectral
representation.

\begin{figure*}[hbpt!]
\begin{center}

\includegraphics[scale=0.5,trim=0cm 0cm 0cm -1cm]{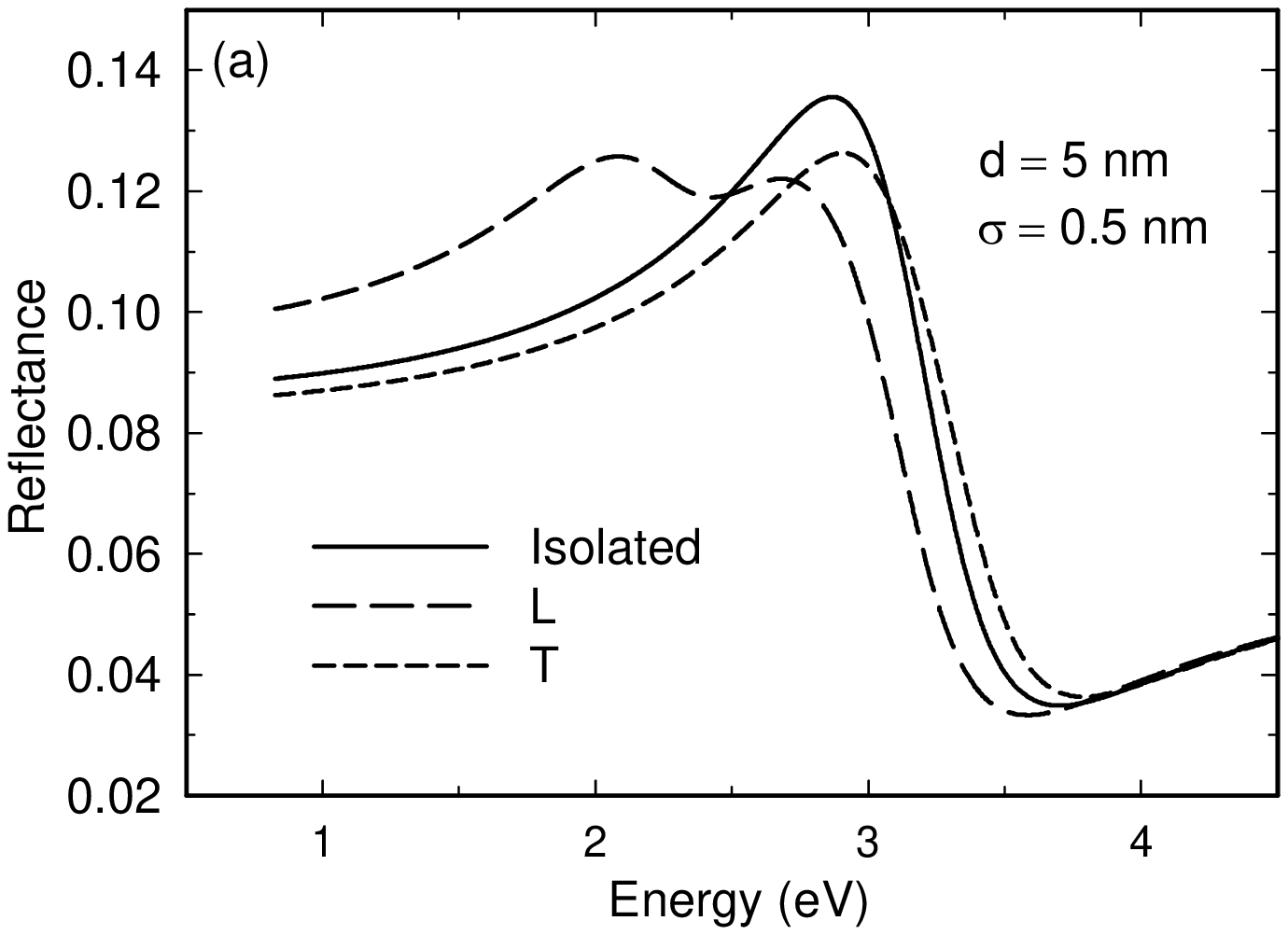}
\includegraphics[scale=0.5,trim=0cm 0cm 0cm -1cm]{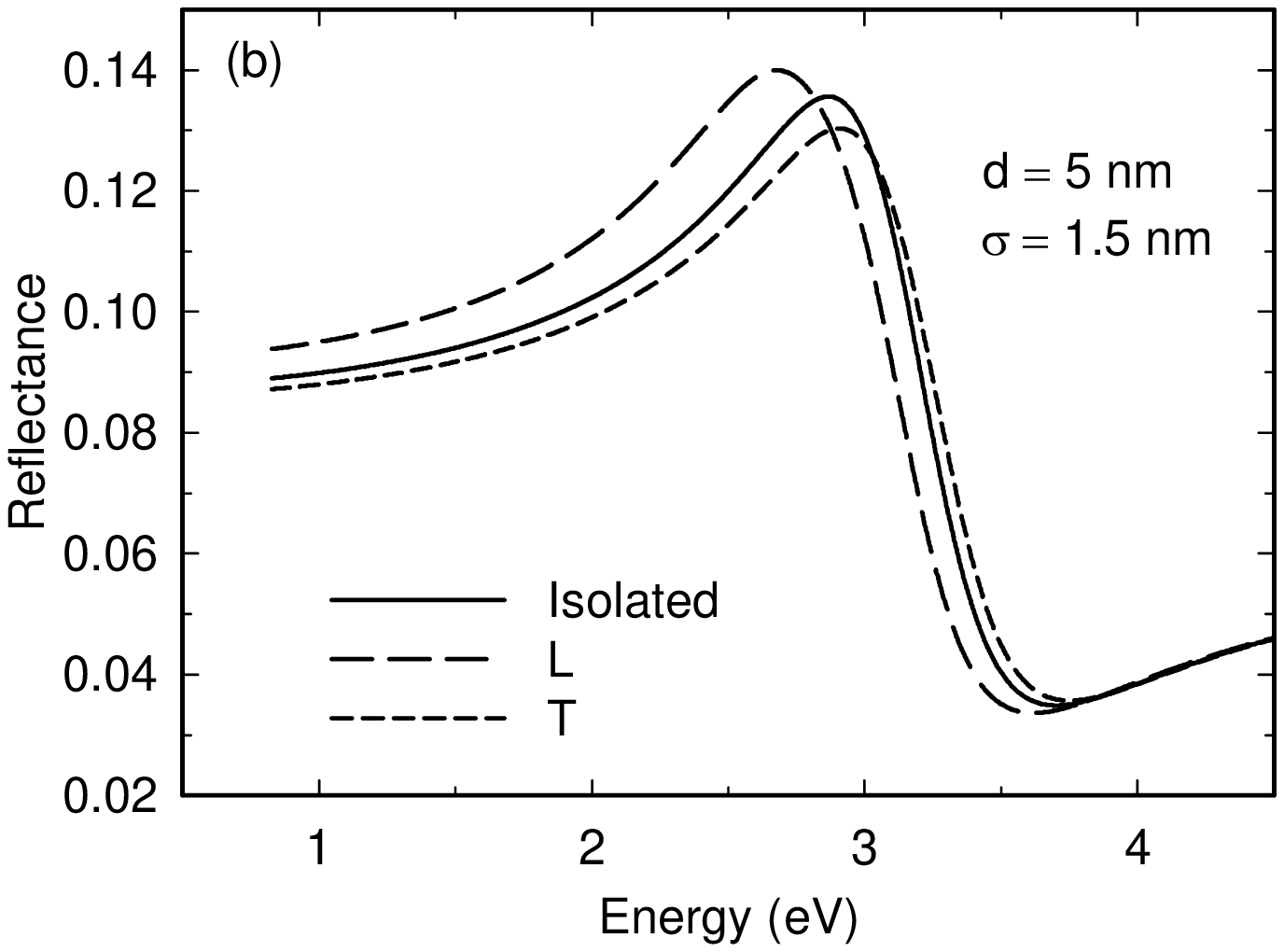}
\includegraphics[scale=0.5,trim=0cm 0cm 0cm -1cm]{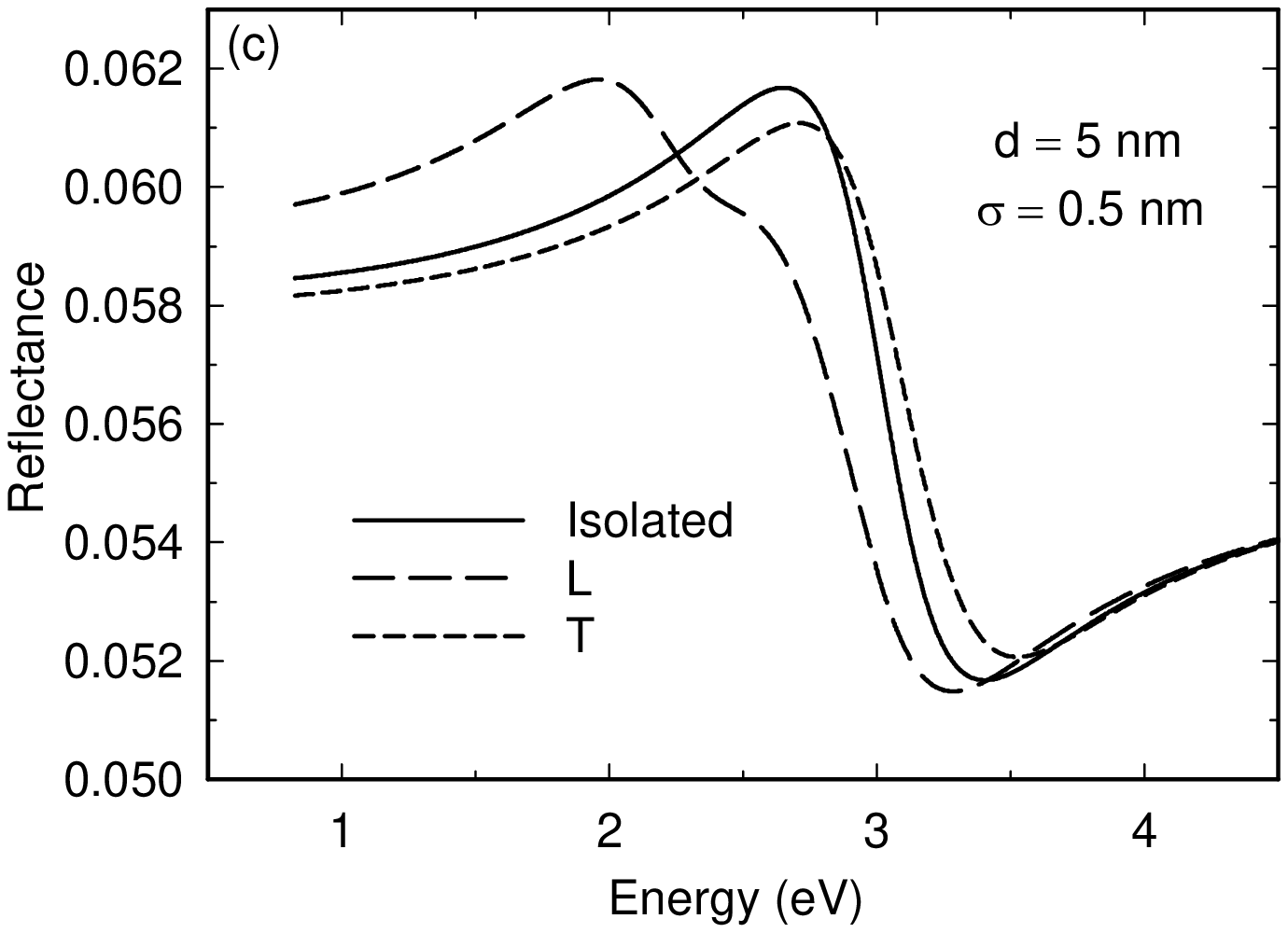}
\includegraphics[scale=0.5,trim=0cm 0cm 0cm -1cm]{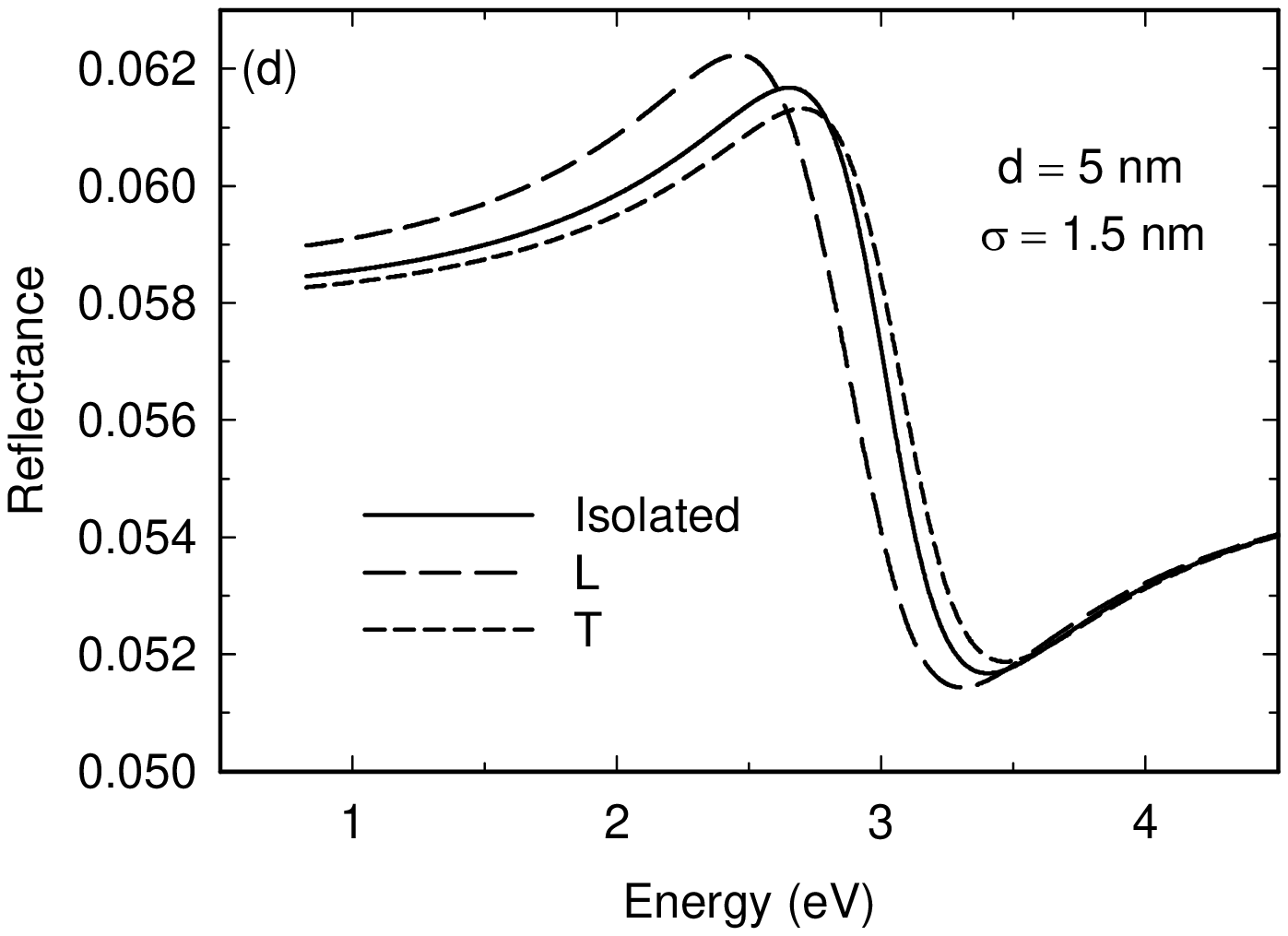}
\end{center}
\caption{}\label{fig:smp1_ref}
\end{figure*}

\textbf{Figure \ref{fig:smp1_ref}}: Normal incidence reflectivity
spectra for an array of dimer with particle diameter $d=5\,$nm for
different spacing $\sigma$, taking volume fraction $p=0.1$ [(a)
and (b)] and $p=0.01$ [(c) and (d)] respectively.

\begin{figure*}[hbpt!]
\begin{center}
\includegraphics[scale=0.5]{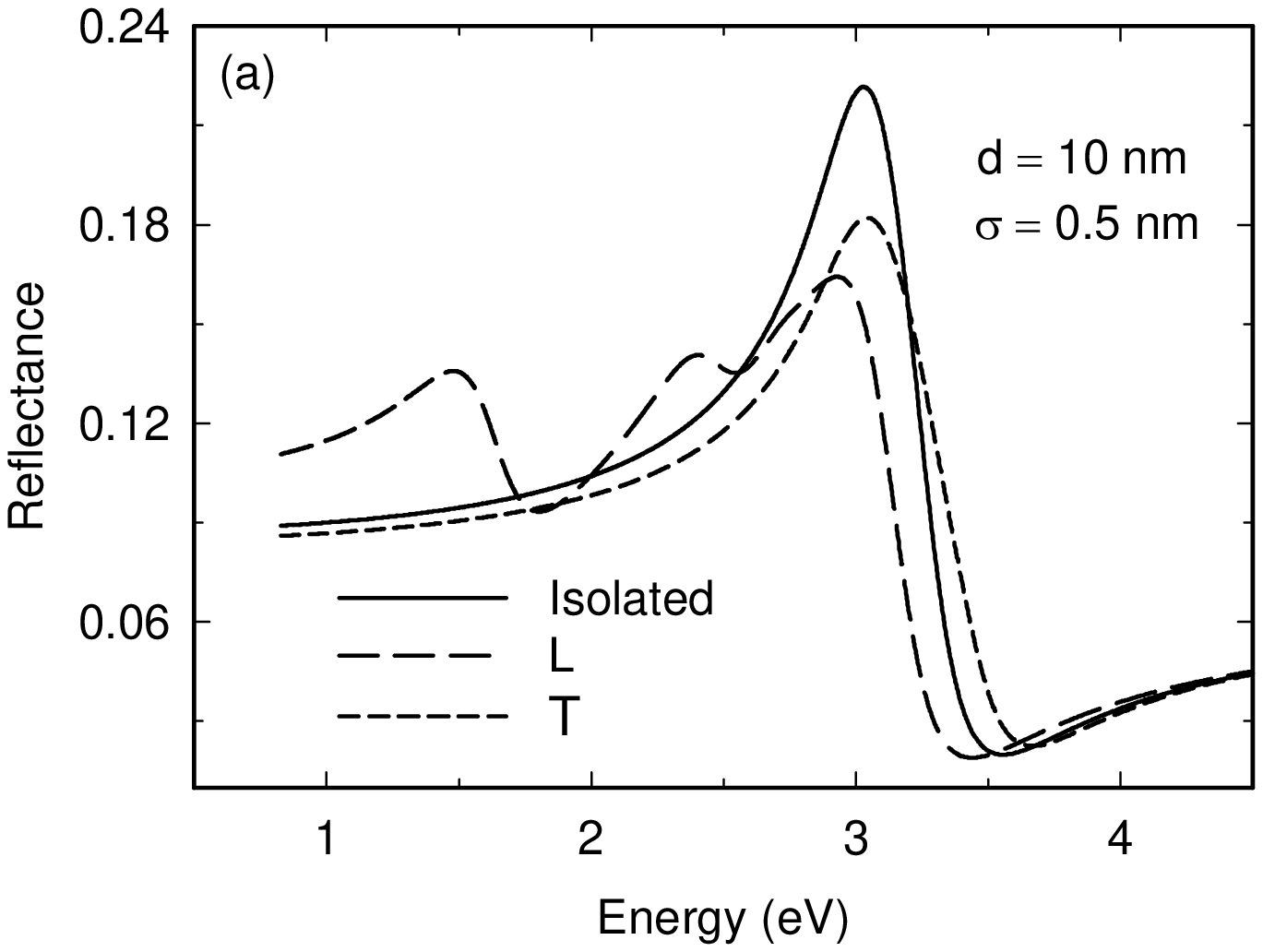}
\includegraphics[scale=0.5]{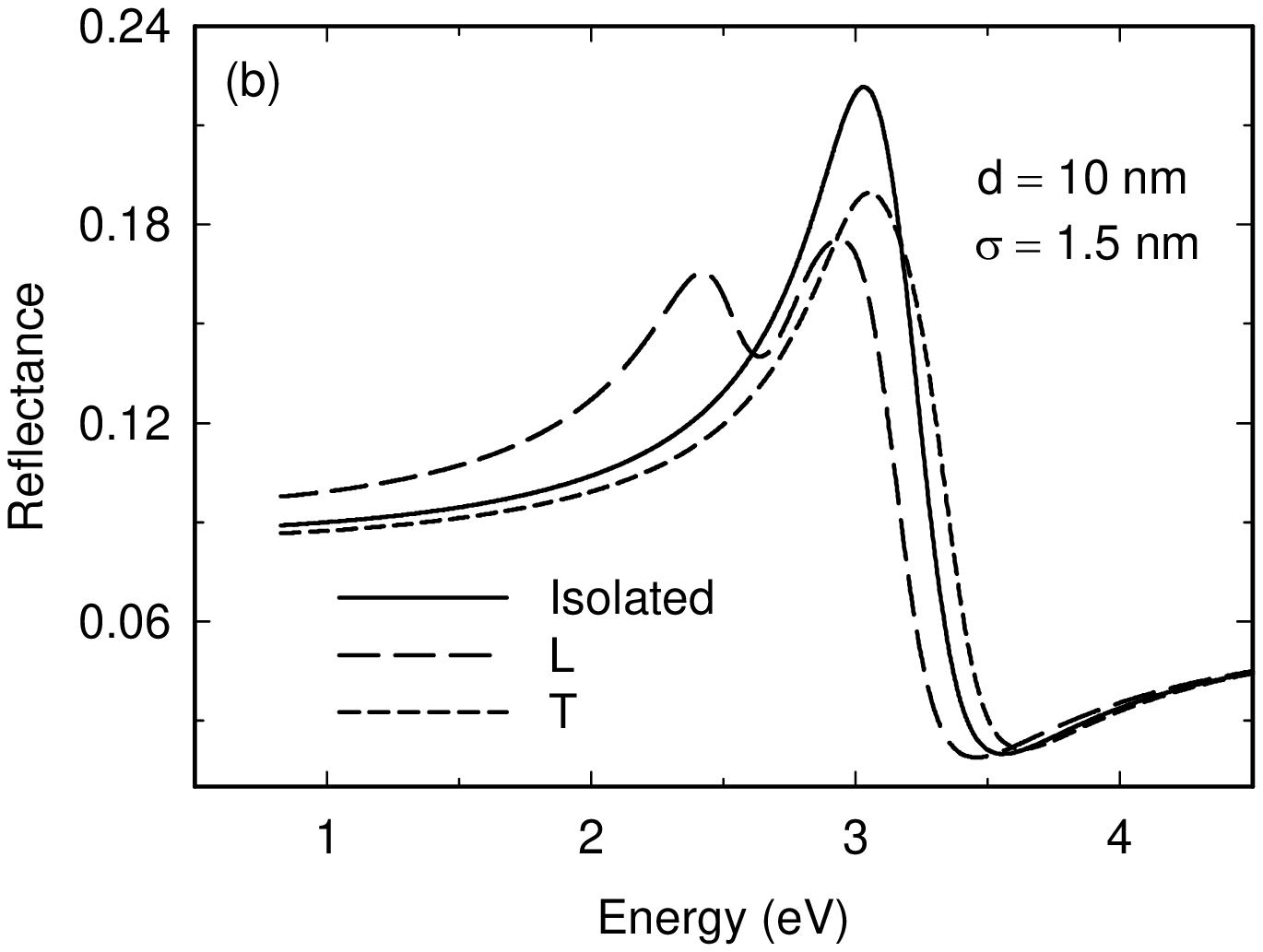}
\includegraphics[scale=0.5,trim=0cm 0cm 0cm -1cm]{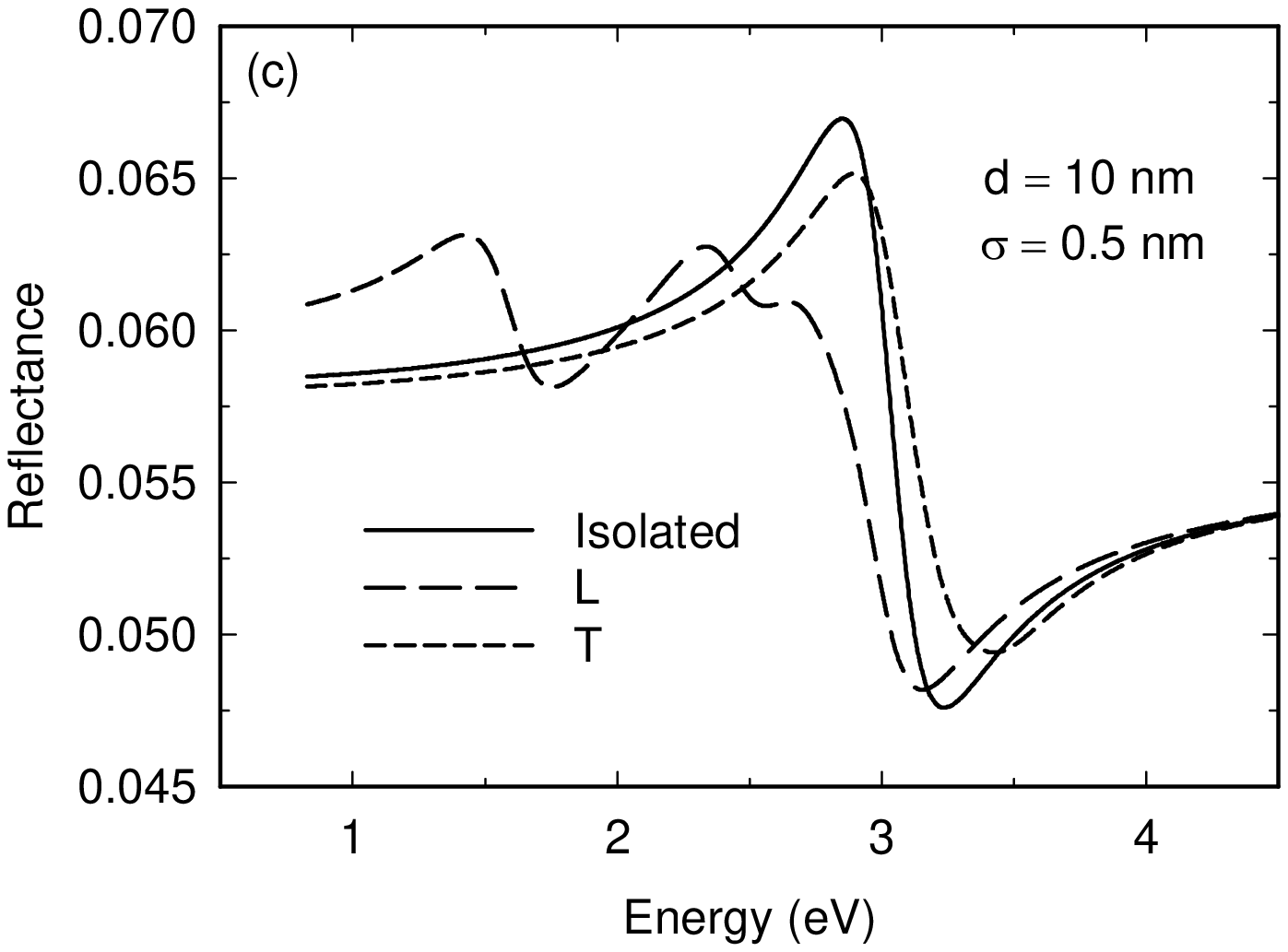}
\includegraphics[scale=0.5,trim=0cm 0cm 0cm -1cm]{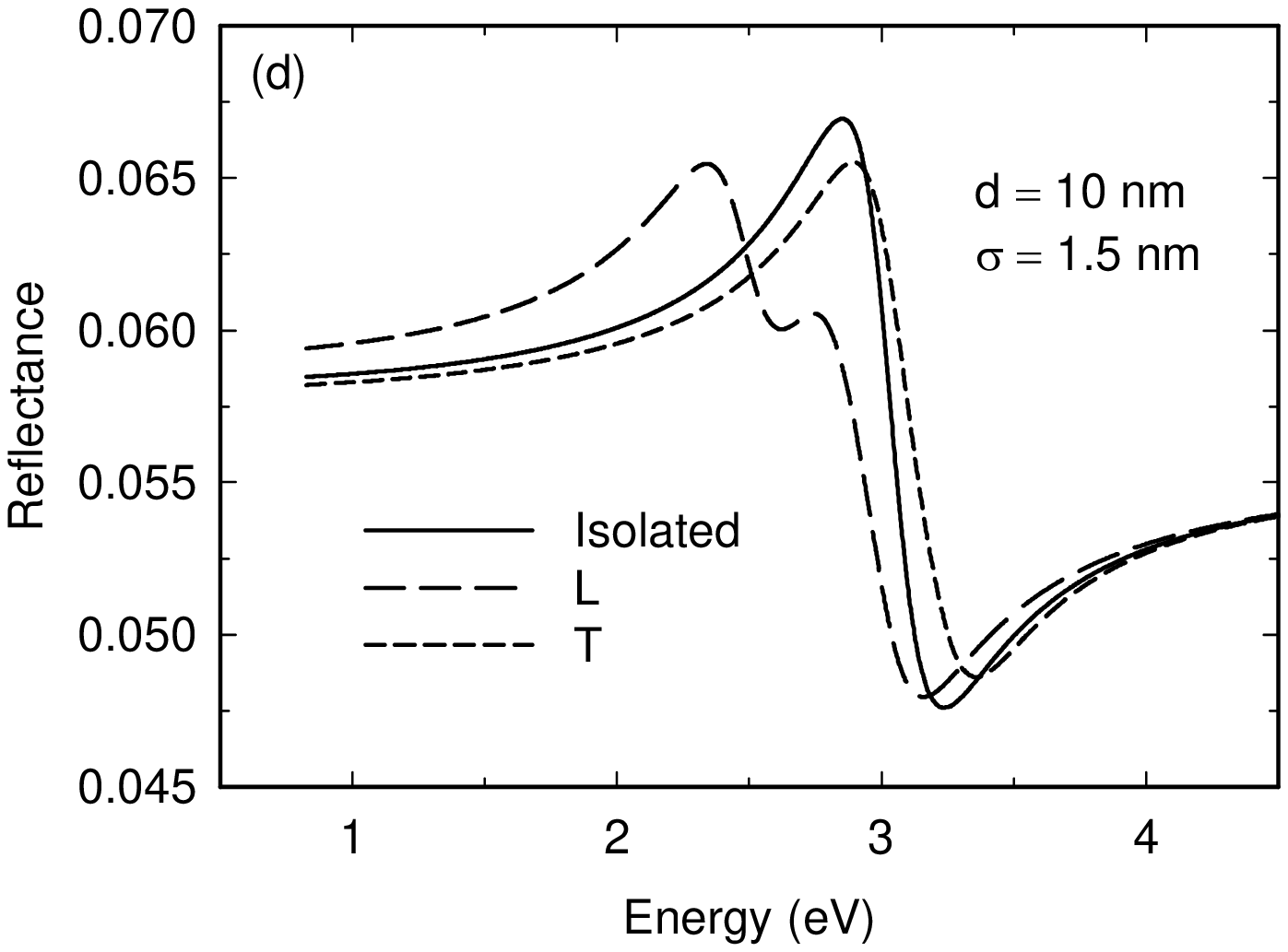}
\end{center}
\caption{}\label{fig:smp2_ref}
\end{figure*}

\textbf{Figure \ref{fig:smp2_ref}}: Normal incidence reflectivity
spectra for an array of dimer with particle diameter $d=10\,$nm.
Others the same as in Fig.~\ref{fig:smp1_ref}.

\end{document}